\documentclass{article}
\usepackage{spconf,amsmath,amssymb,graphicx,cite,hyperref}

\title{AlignBeat: A Latent Variable Model for Multi-Class Beat Tracking from Partially Labeled Data}
\name{Jaehoon Ahn\sthanks{These authors contributed equally.},
Tae Gum Hwang\footnotemark[1],
Moon-Ryul Jung\sthanks{Corresponding author: moon@sogang.ac.kr}}
\address{Sogang University, Seoul, Republic of Korea}
\begin{document}
\ninept
\maketitle
\begin{abstract}
Recent neural beat trackers predict beats and downbeats with two independent frame-wise binary classification heads, a convention adopted because a single multi-class head cannot fully exploit datasets that annotate only beats. The heads can disagree, so Beat This moves each predicted downbeat to the nearest predicted beat. We instead predict a sparse set of grid points, each with an event time and one distribution over downbeat, beat and no event, so a downbeat is a beat by construction and no reconciliation is needed. The alignment between grid points and annotated events is latent, and we fit the model by expectation-maximization. Where no downbeat labels were recorded the event class is latent too and is marginalized out. Over eight-fold cross-validation on eighteen datasets, retaining rather than discarding the beat-only labeled data raises beat CMLt by 4.6 points on the datasets that provide it; downbeat CMLt rises 3.6 points over Beat This, with both F-measures moving by less than one point and no post-processing at any stage.
\end{abstract}
\begin{figure*}[!t]
  \centering
  \includegraphics[width=\textwidth,trim=17bp 4bp 24bp 7bp,clip]{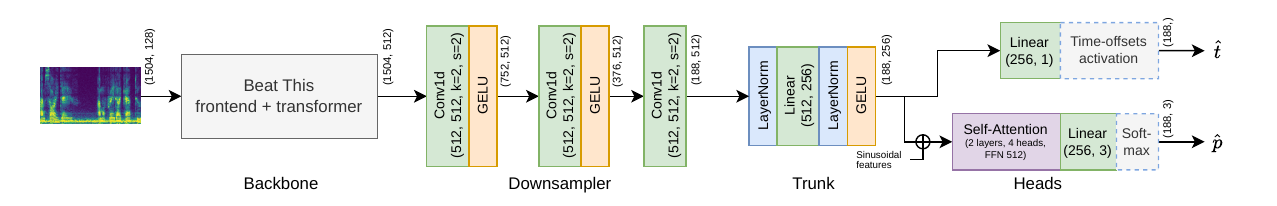}
  \caption{The AlignBeat model. The Beat This frontend and transformer are unchanged; three strided convolutions reduce the $1504$-frame sequence to $N = 188$ grid points, and a shared trunk feeds a time head and a class head. The offset bound $\alpha < 1/2$ makes $\hat{t}$ strictly increasing in $j$. Self-attention over grid points affects the class head only; the time head reads the trunk output directly. Shapes omit the batch dimension.}
  \label{fig:model}
\end{figure*}

\section{Introduction}
Beat tracking locates the times at which a listener would tap to a song; downbeat tracking asks which of those beats begin a bar. The two are structurally coupled; every downbeat is a beat, and a meter together with one bar position determines every subsequent downbeat. Yet some datasets mark every beat as beat or downbeat, and others record beat times alone. Any system trained on their union must confront fragments where the downbeat label was never recorded.

Neural beat trackers predict per-frame activations---at each frame, how likely a beat occurs there and how likely a downbeat does---and recovering the beat and downbeat times themselves from them has conventionally been left to a post-processor. The post-processing has been handled by a dynamic Bayesian network (DBN), known as the bar-pointer model, which represents bar position and tempo as hidden states and recovers them by Viterbi decoding~\cite{whiteley2006bayesian,krebs2013rhythmic,krebs2015efficient}. It has been the standard post-processor since~\cite{bock2016joint}, still used by transformer-based systems a decade later~\cite{zhao2022beat,hung2022modeling} and largely unchanged in the reference implementation~\cite{bock2016madmom}, even though its transition and observation models are hand-specified rather than learned. Removing it has accordingly become a goal in itself~\cite{chen2022toward}, and Beat This~\cite{foscarin2024beat} showed the stage is no longer needed for accurate detection, reaching state-of-the-art F-measure with peak-picking.

Removing the post-processor did not change what the network itself predicts. The three-way output of~\cite{bock2016joint} classified each frame as beat, downbeat or neither, and datasets with beat but no downbeat annotations were simply left out of training. B\"ock and Davies~\cite{bock2020deconstruct} introduced two independent binary heads in order to train on those datasets, reasoning that a multi-class network output cannot fully exploit a dataset in which only one of beat and downbeat is annotated. This two-head design has since become the consensus, at two costs. The DBN of~\cite{bock2016joint} requires the probability of each of the three cases at every frame, which a three-way output supplies directly; two independent activations do not, and the three quantities have to be reconstructed from the two. The two heads are also free to disagree, placing a downbeat where no beat was detected, which Beat This repairs after the fact by moving each predicted downbeat to its nearest predicted beat and discarding any that then collide. A single distribution over $\{\mathrm{DB},\mathrm{B},\varnothing\}$ cannot represent that disagreement, so no such repair is needed.

The consensus has held through the current cycle. Two recent systems keep the two heads~\cite{bolt2026beat,ru2026toward}, and one of them returns to excluding datasets with beat-only annotations from training altogether~\cite{bolt2026beat}. Foscarin et al.~\cite{foscarin2026masked} consider collapsing the two heads back into a single multi-class prediction and set it aside on the same ground, observing that it would be difficult to train on datasets lacking downbeat annotations and would require a new loss.

A single categorical shares one normalization across its three classes: because $\hat p(\mathrm{DB}) + \hat p(\mathrm{B}) + \hat p(\varnothing) = 1$, the loss cannot leave the downbeat probability unconstrained on a beat-only frame without also constraining the other two, unlike a separate downbeat head, whose loss term can simply be masked out on that frame with no effect on the beat head's own prediction. A beat-only annotation is nonetheless informative about such an output. It records that an event occurs at that time, so the loss should reward the model for placing high combined probability on the two event classes, $\hat p(\mathrm{DB}) + \hat p(\mathrm{B})$.
What it does not record is which of the two the event is, so the loss must not push the model toward either. Training the event as an ordinary beat would teach the model that a downbeat there is wrong, which the annotation never states. Marginalizing the class asserts exactly what was annotated and nothing more, resolving the difficulty that Foscarin et al.~\cite{foscarin2026masked} raised but did not address.

In this paper, we introduce AlignBeat, which predicts a sparse set of grid points, as in grid-based object detection~\cite{redmon2017yolo9000}, rather than a dense frame grid, and assigns each grid point an event time and a single three-way class. We treat the alignment between grid points and annotated events as a latent variable and fit the model by expectation--maximization~\cite{dempster1977maximum}, marginalizing the class where no downbeat labels were recorded. Across $8$-fold cross-validation over $18$ datasets, the downbeat continuity metrics (CMLt and AMLt) improve consistently over Beat This.

Our contributions are:
\begin{itemize}
\item A one-stage multi-class formulation in which each grid point carries an event time and one distribution over downbeat, beat and no event, so a downbeat is a beat by construction and no reconciliation step is required.
\item A latent-variable objective in which the alignment between annotated events and grid points is inferred by dynamic programming, and the event class is marginalized where downbeat labels are absent, so fully and beat-only labeled data train one model.
\item An evaluation over $18$ datasets showing $3.6$ points of downbeat CMLt over Beat This, and an ablation showing that using the beat-only labeled datasets rather than discarding them is worth $4.6$ points of beat CMLt on them.
\end{itemize}

\section{Method}

\subsection{Model and latent structure}\label{sec:model}

Any given music fragment carries $M$ annotated event times $t_1 < \dots < t_M$. Where the annotation distinguishes downbeats, each event also carries an observed class $c_i \in \{\mathrm{DB}, \mathrm{B}\}$; where it does not, the $c_i$ are latent and the fragment constrains only that each $t_i$ is an event. Both the alignment cost in~\eqref{eq:match} and the class term in~\eqref{eq:cls} branch on which of the two cases holds.

We keep Beat This's frontend and transformer unchanged and replace its two binary activation functions over the frame grid with a grid point output. Three stacked strided convolutions, each of kernel size and stride $2$, reduce the final sequence to $N$ grid points. Every grid point feature then passes through a shared trunk, from which a class head and a time head read, as is standard in object detection~\cite{redmon2017yolo9000}. The class head emits a distribution $\hat{p}_j$ over $\{\mathrm{DB}, \mathrm{B}, \varnothing\}$, and the time head emits a feature $r_j$ used to produce $\hat{t}_j$, the event's time. Each grid point is therefore computed from consecutive encoder positions, and its predicted time falls inside the interval within the fragment that those positions cover,
\begin{equation}
    \hat{t}_j = \frac{\,j + \tfrac{1}{2} + \alpha \tanh(r_j)\,}{N} , \qquad \alpha = 0.45 < \tfrac{1}{2},
\end{equation}
where $r_j$ is the time head's output and $(j + \tfrac{1}{2})/N$ is the center of that interval. Consecutive centers are $1/N$ apart, and since $\alpha\tanh(r_j)\in(-\alpha,\alpha)$ never closes that gap, $\hat{t}_j$ stays within grid point $j$'s own cell and $\hat{t}_0 < \dots < \hat{t}_{N-1}$ is guaranteed for every $r$. The time head's output layer is initialized with zero weights and bias, so $r_j = 0$ at the start of training and every grid point begins at its own center. In the implementation $\hat{t}$ is rescaled by $1504/1500$ so the padded grid maps onto the fragment. The class head begins with two transformer layers in which grid points attend to one another, so that a grid point can see whether a neighbor is already claiming the same event. We use $N = 188$, three halvings of the padded $1504$-frame fragment, which supports up to $188$ events per $30$-second fragment ($376$~bpm). Each annotated event needs a grid point of its own, so $N \geq M$ must hold; five tracks contain a denser window than that, and fragments drawn from them are dropped.

\subsection{Expectation--maximization}

A frame-wise tracker predicts, at each frame $t$, whether a beat or downbeat occurs there; the target is aligned with the output by construction, since both are indexed by frame. Our model instead places $N$ grid points over the fragment and predicts at each one an event time and an event class. Nothing indexes the $M$ annotated events to the $N$ grid points, so before any loss can be computed we must decide which grid point is responsible for which annotated event. We write that decision as an alignment $\sigma$, with $\sigma(i) = j$ meaning that event $i$ is predicted by grid point $j$. It is never annotated, so we treat it as a latent variable. For beat-only labeled data the event class is unobserved as well.
We fit the model by expectation--maximization: the E-step selects, at the current parameters $\theta$, the single lowest-cost alignment rather than a distribution over alignments, and the M-step takes a gradient step on the loss evaluated at that alignment, with the class of a beat-only labeled fragment marginalized out.
 We score every pairing of an event $i$ with a grid point $j$ by a time term and a class term,

\begin{equation}
    \begin{aligned}
        \mathcal{L}_{\mathrm{match}}(i, j)
        &= \lambda \lvert t_i - \hat{t}_j \rvert \\
        &\quad + \begin{cases}
            -\log \hat{p}_j(c_i) & \text{fully labeled} \\
            -\log\!\big(\hat{p}_j(\mathrm{DB}) + \hat{p}_j(\mathrm{B})\big) & \text{beat-only labeled.}
        \end{cases}
    \end{aligned}
    \label{eq:match}
\end{equation}

The time term is the negative log-likelihood of a Laplace density of fixed scale $b$ centered on $\hat{t}_j$; its constant term carries no gradient and is dropped, leaving $\lambda \lvert t_i - \hat{t}_j \rvert$ with $\lambda = 1/b$. We set $b = 35$~ms, half the F-measure tolerance~\cite{davies2009evaluation}, giving $\lambda = 857$ as a fraction of the $30$-second fragment.

For fully labeled data the pairing is charged the negative log-likelihood of the recorded class. For beat-only labeled data the annotation asserts only that an event occurs at $t_i$, and the grid point's belief in that is its combined probability on the two event classes, $\hat{p}_j(\mathrm{DB}) + \hat{p}_j(\mathrm{B})$.

\textbf{E-step.} We write $\hat{\sigma}$ for the assignment of events to distinct grid points that minimizes the total cost $\sum_i \mathcal{L}_{\mathrm{match}}(i, \hat{\sigma}(i))$. Only assignments in which later events match later grid points are admitted, since both sequences are sorted in time. Writing $D(i,j)$ for the cost of matching the first $i$ events within the first $j$ grid points, this minimum satisfies
\begin{equation}
    D(i, j) = \min\big\{\, D(i, j-1),\; D(i-1, j-1) + \mathcal{L}_{\mathrm{match}}(i, j) \,\big\},
\end{equation}
with $D(0, j) = 0$ and $D(i, 0) = \infty$ for $i \geq 1$. The optimum is $D(M,N)$, and $\hat{\sigma}$ is recovered by dynamic programming in $O(MN)$. A solution exists whenever $N \geq M$, which the grid-point count of Section~\ref{sec:model} satisfies.
The recursion is evaluated without gradient, and $\hat\sigma$ enters the M-step as a constant. Because $\hat\sigma$ is the single lowest-cost alignment rather than a posterior over alignments, this is a hard-EM (Viterbi-style) E-step: the M-step ascends the joint likelihood of the data and $\hat\sigma$ rather than the exact marginal likelihood obtained by summing over all alignments, so the usual EM guarantee that each step cannot decrease the marginal likelihood does not transfer directly.

\textbf{M-step.} With the alignment frozen, we take a gradient step on a loss with three terms: a class term, a time term on matched pairs, and a background term on the grid points left unmatched. The class term is an ordinary cross-entropy for fully labeled data, and the marginal likelihood of an event for beat-only labeled data,

\begin{equation}
    \mathrm{cls}(\theta) =
    \begin{cases}
        -\sum_i \log \hat{p}_{\hat{\sigma}(i)}(c_i) & \text{fully labeled} \\[4pt]
        -\sum_i \log\!\big(\hat{p}_{\hat{\sigma}(i)}(\mathrm{DB}) + \hat{p}_{\hat{\sigma}(i)}(\mathrm{B})\big) & \text{beat-only labeled.}
    \end{cases}
    \label{eq:cls}
\end{equation}

The second branch is the quantity the alignment cost already charges an unlabeled pair, so the matching and the loss agree on what such an annotation asserts. The fragment's loss adds the timing error on matched pairs and a background term on every grid point the alignment did not use,

\begin{equation}
    \mathcal{L}(\theta) = \mathrm{cls}(\theta) + \lambda \sum_{i} \big\lvert t_i - \hat{t}_{\hat{\sigma}(i)} \big\rvert - \gamma \!\!\sum_{j \notin \mathrm{im}(\hat{\sigma})} \!\! \log \hat{p}_j(\varnothing) ,
\end{equation}

where $\mathrm{im}(\hat{\sigma})$ is the set of grid points the alignment uses, $\gamma = 1$, and a gradient step on $\mathcal{L}$ completes the iteration.

\subsection{Inference}

Pieces longer than the $1500$-frame training fragment are covered by overlapping fragments, discarding a $\beta = 8$-frame margin, one grid-point spacing, at each internal boundary so every event is decoded away from a boundary; the last fragment is slid back to end at the final frame rather than zero-padded. A piece shorter than one fragment is zero-padded instead. Within a fragment, grid point $j$ is retained when
\begin{equation}
    1 - \hat{p}_j(\varnothing) \geq \tau ,
    \label{eq:tau}
\end{equation}
with $\tau = 0.5$, and is then labeled by the larger of $\hat{p}_j(\mathrm{DB})$ and $\hat{p}_j(\mathrm{B})$.

\section{Experimental setup}

We evaluate on the $18$ datasets distributed with Beat This~\cite{foscarin2024beat} under its $8$-fold split~\cite{nieto2019harmonix,goto2002rwc,goto2006aist,driedger2019towards,eremenko2018audio,foster2021filosax,foscarin2020asap,gillick2019learning,xi2018guitarset,krebs2013rhythmic,fujinaga2012one,hainsworth2004particle,gouyon2005computational,holzapfel2012selective,rocamora2015audio,davies2009evaluation}, with GTZAN~\cite{marchand2015swing} held out entirely as a test set. Two of the $18$, Simac~\cite{gouyon2005computational} and SMC~\cite{holzapfel2012selective}, carry no downbeat annotations, so beat metrics are averaged over $4556$ pieces and downbeat metrics over $3744$. We report F-measure and the continuity measures CMLt and AMLt~\cite{davies2009evaluation}, computed with \texttt{mir\_eval}~\cite{raffel2014mir_eval} over whole pieces with the first five seconds discarded, and use no DBN or other post-processing anywhere. Our baseline is Beat This's released per-fold checkpoints, scored by us through the identical pipeline; on GTZAN they reproduce the figures quoted in~\cite{foscarin2026masked} to within $0.07$ on every metric.

We initialize the encoder from Beat This's released checkpoint for the corresponding fold, because we reuse its backbone unchanged, and train the full model, encoder and heads together, for a further $20$ epochs. The three models reported on GTZAN are trained the same way on all cross-validation data, differing only in seed.

Fragments are $1500$ frames at $50$~fps, with Beat This's tempo, pitch and masking augmentation unchanged. We train in mixed precision with AdamW~\cite{loshchilov2017decoupled} at $3\times10^{-4}$, weight decay $0.01$, $1000$ warmup steps and a cosine schedule, batch size $8$ with gradient accumulation over $8$ steps.

\begin{table}[!t]
  \centering
  \caption{Beat results per dataset, sorted by CMLt gain. All datasets carry beat annotations. BT denotes Beat This~\cite{foscarin2024beat}.}
  \label{tab:perdatasetbeat}
  \small
  \setlength{\tabcolsep}{3.5pt}
  \begin{tabular}{@{}lr rr rr rr@{}}
    \hline
    & & \multicolumn{2}{c}{F-measure} & \multicolumn{2}{c}{CMLt} & \multicolumn{2}{c}{AMLt} \\
    \cline{3-4}\cline{5-6}\cline{7-8}
    Dataset & $n$ & BT & Ours & BT & Ours & BT & Ours \\
    \hline
    Simac & 595 & 77.9 & \textbf{83.4} & 55.7 & \textbf{73.9} & 85.1 & \textbf{85.6} \\
    GuitarSet & 180 & 92.0 & \textbf{96.1} & 82.4 & \textbf{93.1} & 90.1 & \textbf{94.3} \\
    JAAH & 113 & 95.1 & \textbf{95.5} & 88.5 & \textbf{90.2} & 90.3 & \textbf{91.4} \\
    TapCorrect & 101 & 93.0 & \textbf{93.3} & 81.9 & \textbf{83.5} & 89.0 & \textbf{90.3} \\
    ASAP & 473 & \textbf{76.3} & 76.1 & 50.3 & \textbf{51.7} & 57.8 & \textbf{58.8} \\
    RWC RF & 15 & 94.5 & \textbf{95.0} & 87.4 & \textbf{88.6} & \textbf{94.0} & 91.7 \\
    Beatles & 179 & 94.5 & \textbf{94.9} & 87.2 & \textbf{88.2} & 92.9 & \textbf{93.6} \\
    RWC Pop & 100 & 96.1 & \textbf{96.3} & 90.1 & \textbf{90.8} & 93.6 & \textbf{94.7} \\
    Harmonix & 911 & 95.8 & \textbf{95.9} & 89.9 & \textbf{90.2} & 94.0 & \textbf{94.2} \\
    Groove MIDI & 336 & \textbf{93.7} & 93.6 & 87.1 & \textbf{87.2} & 91.4 & \textbf{91.7} \\
    RWC Jazz & 50 & \textbf{83.3} & 82.5 & \textbf{72.4} & \textbf{72.4} & 75.1 & \textbf{75.8} \\
    Filosax & 48 & \textbf{99.5} & \textbf{99.5} & \textbf{98.8} & 98.7 & \textbf{98.8} & 98.7 \\
    HJDB & 235 & \textbf{98.2} & 97.8 & \textbf{97.2} & 97.1 & \textbf{97.9} & 97.7 \\
    Candombe & 35 & \textbf{99.7} & 99.6 & \textbf{99.8} & 99.6 & \textbf{99.8} & 99.6 \\
    Hainsworth & 222 & \textbf{91.9} & 91.3 & \textbf{84.0} & 83.9 & \textbf{90.9} & 90.7 \\
    Ballroom & 685 & \textbf{97.5} & 97.3 & \textbf{96.4} & 96.2 & 97.0 & \textbf{97.5} \\
    SMC & 217 & \textbf{62.7} & 61.5 & \textbf{51.4} & 51.0 & \textbf{61.0} & 59.5 \\
    RWC Classical & 61 & \textbf{77.1} & 76.0 & \textbf{51.8} & 51.2 & \textbf{59.1} & 58.2 \\
    \hline All & 4556 & 89.3 & \textbf{90.0} & 79.2 & \textbf{82.3} & 86.8 & \textbf{87.3} \\
    \hline
  \end{tabular}
\end{table}

\begin{table}[!t]
  \centering
  \caption{Downbeat results per dataset, sorted by CMLt gain. BT denotes Beat This~\cite{foscarin2024beat}.}
  \label{tab:perdatasetdownbeat}
  \small
  \setlength{\tabcolsep}{3.5pt}
  \begin{tabular}{@{}lr rr rr rr@{}}
    \hline
    & & \multicolumn{2}{c}{F-measure} & \multicolumn{2}{c}{CMLt} & \multicolumn{2}{c}{AMLt} \\
    \cline{3-4}\cline{5-6}\cline{7-8}
    Dataset & $n$ & BT & Ours & BT & Ours & BT & Ours \\
    \hline
    GuitarSet & 180 & 88.1 & \textbf{93.6} & 79.8 & \textbf{91.5} & 87.4 & \textbf{91.9} \\
    TapCorrect & 101 & 86.4 & \textbf{87.9} & 68.2 & \textbf{76.4} & 76.5 & \textbf{81.8} \\
    JAAH & 113 & \textbf{85.0} & 84.9 & 71.5 & \textbf{77.8} & 74.2 & \textbf{78.7} \\
    Hainsworth & 222 & \textbf{80.0} & 78.6 & 63.6 & \textbf{69.1} & 75.1 & \textbf{77.5} \\
    ASAP & 473 & \textbf{61.2} & 59.9 & 23.5 & \textbf{28.6} & \textbf{43.9} & 42.8 \\
    RWC Jazz & 50 & \textbf{80.7} & \textbf{80.7} & 71.3 & \textbf{75.9} & 76.7 & \textbf{78.1} \\
    RWC Classical & 61 & \textbf{66.3} & 65.7 & 34.2 & \textbf{38.8} & 47.6 & \textbf{51.1} \\
    Beatles & 179 & 88.8 & \textbf{89.7} & 73.7 & \textbf{77.9} & 82.0 & \textbf{84.5} \\
    RWC RF & 15 & 91.9 & \textbf{92.6} & 82.2 & \textbf{85.7} & 82.2 & \textbf{85.7} \\
    Harmonix & 911 & 90.7 & \textbf{91.1} & 81.2 & \textbf{84.6} & 85.9 & \textbf{89.2} \\
    Groove MIDI & 336 & \textbf{82.1} & 81.9 & 72.7 & \textbf{74.8} & 81.9 & \textbf{84.0} \\
    RWC Pop & 100 & 93.7 & \textbf{94.1} & 87.1 & \textbf{89.1} & 89.2 & \textbf{92.2} \\
    Filosax & 48 & \textbf{98.5} & \textbf{98.5} & 96.1 & \textbf{97.3} & 96.1 & \textbf{97.3} \\
    Ballroom & 685 & \textbf{95.3} & 95.0 & 92.9 & \textbf{93.8} & 94.1 & \textbf{95.5} \\
    HJDB & 235 & \textbf{96.6} & 96.0 & 94.8 & \textbf{95.3} & 95.6 & \textbf{96.1} \\
    Candombe & 35 & \textbf{99.7} & 99.6 & \textbf{99.9} & 99.8 & \textbf{99.9} & 99.8 \\
    \hline All & 3744 & 86.0 & \textbf{86.1} & 73.7 & \textbf{77.2} & 80.6 & \textbf{82.7} \\
    \hline
  \end{tabular}
\end{table}

\section{Results}\label{sec:results}

Tables~\ref{tab:perdatasetbeat} and~\ref{tab:perdatasetdownbeat} give the cross-validation results.

On pieces where Beat This tracks beats at double or half tempo, our model more often tracks beats at the correct metrical level and phase. The evidence is that over all $4556$ pieces beat CMLt rises by $3.1$ points while AMLt, which also accepts double and half tempo and an offbeat phase, rises by only $0.5$. The effect is largest where the two metrics were furthest apart: on Simac beat CMLt rises from $55.7$ to $73.9$ against an AMLt already at $85.1$, and on GuitarSet CMLt goes from $82.4$ to $93.1$ against AMLt's $90.1$ to $94.3$.

Beat and downbeat F-measure scores themselves are largely unchanged between Beat This and AlignBeat. Averaged over all $4556$ pieces, beat F-measure rises by $0.7$ points and downbeat F-measure by $0.1$, and no dataset loses more than $1.2$ beat or $1.4$ downbeat F-measure. Among the datasets, the beat F-measure scores for Simac and GuitarSet increased the most, by $5.5$ and $4.1$ points respectively; on the remaining sixteen datasets, every difference is within $1.2$ points.

Downbeat CMLt improves on $15$ of the $16$ datasets that have downbeat annotations, losing on Candombe. Downbeat F-measure improves or ties on $8$ of $16$; on the remaining eight it never loses by more than two points.

Table~\ref{tab:gtzan} compares tracking metrics on the held-out GTZAN test set. Downbeat CMLt gains $2.8$ points over Beat This and AMLt $3.3$, while both F-measures fall by less than one point and beat continuity stays within $0.3$. MDM~\cite{foscarin2026masked} leads on every metric but beat AMLt, using an ensemble of three models at eight inference steps. BAT~\cite{ru2026toward} injects musical period and phase priors into the attention mechanism.

\begin{table}[!t]
  \centering
  \caption{GTZAN (held-out), $993$ tracks. Rows marked $\dagger$ are quoted; the others are scored by us.}  \label{tab:gtzan}
  \small
  \setlength{\tabcolsep}{3pt}
  \begin{tabular}{@{}l rrr rrr@{}}
    \hline
    & \multicolumn{3}{c}{Beat} & \multicolumn{3}{c}{Downbeat} \\
    \cline{2-4}\cline{5-7}
    & F & CMLt & AMLt & F & CMLt & AMLt \\
    \hline
    Beat This~\cite{foscarin2024beat} & 89.1 & 79.8 & 89.8 & 78.3 & 67.2 & 79.1 \\
    BAT~\cite{ru2026toward}$^\dagger$ & 88.7 & 81.5 & \textbf{93.8} & 74.7 & 67.3 & 85.7 \\
    MDM~\cite{foscarin2026masked}$^\dagger$ & \textbf{89.7} & \textbf{82.9} & 92.5 & \textbf{79.5} & \textbf{76.4} & \textbf{88.5} \\
    AlignBeat (Ours) & 88.8 & 79.5 & 90.1 & 77.5 & 70.0 & 82.4 \\
    \hline
  \end{tabular}
\end{table}

\subsection{Keeping or discarding beat-only labeled data}

We have tested how keeping the beat-only labeled data in training affects beat and downbeat tracking performance. We compare two training conditions: DROP, which excludes the beat-only labeled datasets, Simac~\cite{gouyon2005computational} and SMC~\cite{holzapfel2012selective}, from training entirely; and KEEP, which includes them, supervised on event times with the class marginalized out.

Table~\ref{tab:ablation} compares DROP with KEEP on the $812$ pieces from Simac and SMC and reports the remaining $3744$ pieces as a control. Relative to DROP, KEEP gains $1.8$ beat F-measure and $5.3$ beat CMLt on Simac, and $1.4$ beat F-measure and $2.8$ beat CMLt on SMC. On Simac, beat AMLt is nearly unchanged across the three systems in Table~\ref{tab:ablation}, meaning all three already produce a beat sequence at some tempo and phase that AMLt accepts.

On the remaining $3744$ pieces the two arms agree to within $0.2$ on every metric except downbeat CMLt, which falls by $0.4$. A beat-only annotation states only that an event occurred, not whether it is a beat or a downbeat, so the downbeat class is still learned entirely from the fully labeled datasets.

\begin{table}[!t]
  \centering
  \caption{Beat and downbeat tracking with AlignBeat trained without the beat-only labeled datasets (DROP) or with them (KEEP). Simac ($595$ pieces) and SMC ($217$) carry no downbeat annotations, so downbeat scores exist only for the remaining $3744$.}
  \label{tab:ablation}
  \small
  \setlength{\tabcolsep}{1.4pt}
  \begin{tabular}{@{}l rrr @{\hspace{8pt}}rrr @{\hspace{8pt}}rrr@{}}
    \hline
    & \multicolumn{3}{c}{Beat This~\cite{foscarin2024beat}} & \multicolumn{3}{c}{DROP} & \multicolumn{3}{c}{KEEP} \\
    \cline{2-4}\cline{5-7}\cline{8-10}
    Dataset & F & CMLt & AMLt & F & CMLt & AMLt & F & CMLt & AMLt \\
    \hline
    \multicolumn{10}{@{}l}{\itshape Beat} \\
    Simac & 77.9 & 55.7 & 85.1 & 81.6 & 68.6 & 85.5 & \textbf{83.4} & \textbf{73.9} & \textbf{85.6} \\
    SMC & \textbf{62.7} & \textbf{51.4} & \textbf{61.0} & 60.1 & 48.2 & 58.5 & 61.5 & 51.0 & 59.5 \\
    Others & 92.6 & 84.5 & 88.6 & \textbf{92.8} & \textbf{85.6} & 89.1 & 92.7 & 85.5 & \textbf{89.2} \\
    \hline
    \multicolumn{10}{@{}l}{\itshape Downbeat} \\
    Others & 86.0 & 73.7 & 80.6 & \textbf{86.2} & \textbf{77.6} & \textbf{82.7} & 86.1 & 77.2 & \textbf{82.7} \\
    \hline
  \end{tabular}
\end{table}

\section{Conclusion}

We have presented AlignBeat, which replaces the two independent frame-wise binary activations of current beat trackers with a single categorical distribution over downbeat, beat and no event, carried on a sparse set of grid points that each predict a continuous time. Treating the alignment between grid points and annotated events as a latent variable, and the event class as latent as well where downbeat labels are absent, makes that representation trainable by expectation--maximization on fully and partially labeled data together.

Our model restores the multi-class output abandoned in~\cite{bock2020deconstruct} and declined again in~\cite{foscarin2026masked}. Across $8$-fold cross-validation downbeat CMLt gains $3.6$ points and AMLt $2.1$, and beat CMLt gains $3.1$ against an AMLt gain of only $0.5$, so the predicted events move to the annotated metrical level rather than simply becoming more numerous. Both F-measures move by less than one point, so nothing is traded away for it.

Simac and SMC could previously only be discarded~\cite{bolt2026beat} or masked; when these datasets are included in the training data, beat CMLt on them rises $4.6$ points over the same model trained without them, and Simac gains $18.2$ points on beat CMLt over Beat This. The benefit of the beat-only data does not reach the fully labeled datasets, where the two training conditions agree to within $0.4$ on every metric. Treating the meter of a piece as a further latent variable may allow it to, and we also plan to extend the latent-variable formulation to fully unlabeled audio.

\bibliographystyle{IEEEbib}
\bibliography{strings,refs}

@inproceedings{foscarin2024beat,
    author = {Francesco Foscarin and Jan Schl{\"u}ter and Gerhard Widmer},
    title = {Beat this! Accurate beat tracking without {DBN} postprocessing},
    year = 2024,
    month = nov,
    booktitle = {Proceedings of the 25th International Society for Music Information Retrieval Conference (ISMIR)},
    address = {San Francisco, CA, United States},
}

@article{foscarin2026masked,
  title={Masked diffusion enables coherent beat tracking},
  author={Foscarin, Francesco and Korzeniowski, Filip and Vogl, Richard},
  journal={arXiv preprint arXiv:2608.04624},
  year={2026}
}

@inproceedings{whiteley2006bayesian,
  title={Bayesian Modelling of Temporal Structure in Musical Audio.},
  author={Whiteley, Nick and Cemgil, Ali Taylan and Godsill, Simon J},
  booktitle={ISMIR},
  pages={29--34},
  year={2006}
}

@inproceedings{krebs2015efficient,
  title={An Efficient State-Space Model for Joint Tempo and Meter Tracking.},
  author={Krebs, Florian and B{\"o}ck, Sebastian and Widmer, Gerhard},
  booktitle={ISMIR},
  pages={72--78},
  year={2015}
}

@inproceedings{bock2020deconstruct,
  title={Deconstruct, Analyse, Reconstruct: How to improve Tempo, Beat, and Downbeat Estimation.},
  author={B{\"o}ck, Sebastian and Davies, Matthew EP},
  booktitle={ISMIR},
  pages={574--582},
  year={2020}
}

@inproceedings{bock2016joint,
  title={Joint Beat and Downbeat Tracking with Recurrent Neural Networks.},
  author={B{\"o}ck, Sebastian and Krebs, Florian and Widmer, Gerhard},
  booktitle={ISMIR},
  pages={255--261},
  year={2016},
  organization={New York City}
}

@inproceedings{bock2016madmom,
  title={Madmom: A new python audio and music signal processing library},
  author={B{\"o}ck, Sebastian and Korzeniowski, Filip and Schl{\"u}ter, Jan and Krebs, Florian and Widmer, Gerhard},
  booktitle={Proceedings of the 24th ACM international conference on Multimedia},
  pages={1174--1178},
  year={2016}
}

@inproceedings{chen2022toward,
  title={Toward postprocessing-free neural networks for joint beat and downbeat estimation.},
  author={Chen, Tsung-Ping and Su, Li},
  booktitle={ISMIR},
  pages={27--35},
  year={2022}
}

@article{zhao2022beat,
  title={Beat transformer: Demixed beat and downbeat tracking with dilated self-attention},
  author={Zhao, Jingwei and Xia, Gus and Wang, Ye},
  journal={arXiv preprint arXiv:2209.07140},
  year={2022}
}

@inproceedings{hung2022modeling,
  title={Modeling beats and downbeats with a time-frequency transformer},
  author={Hung, Yun-Ning and Wang, Ju-Chiang and Song, Xuchen and Lu, Wei-Tsung and Won, Minz},
  booktitle={ICASSP 2022-2022 IEEE International Conference on Acoustics, Speech and Signal Processing (ICASSP)},
  pages={401--405},
  year={2022},
  organization={IEEE}
}

@article{dempster1977maximum,
  title={Maximum likelihood from incomplete data via the EM algorithm},
  author={Dempster, Arthur P and Laird, Nan M and Rubin, Donald B},
  journal={Journal of the royal statistical society: series B (methodological)},
  volume={39},
  number={1},
  pages={1--22},
  year={1977},
  publisher={Wiley Online Library}
}

@article{davies2009evaluation,
  title={Evaluation methods for musical audio beat tracking algorithms},
  author={Davies, Matthew EP and Degara, Norberto and Plumbley, Mark D},
  journal={Queen Mary University of London, Centre for Digital Music, Tech. Rep. C4DM-TR-09-06},
  year={2009}
}

@inproceedings{nieto2019harmonix,
  title={The Harmonix Set: Beats, Downbeats, and Functional Segment Annotations of Western Popular Music.},
  author={Nieto, Oriol and McCallum, Matthew C and Davies, Matthew EP and Robertson, Andrew and Stark, Adam M and Egozy, Eran},
  booktitle={ISMIR},
  pages={565--572},
  year={2019}
}

@inproceedings{goto2002rwc,
  title={RWC Music Database: Popular, Classical and Jazz Music Databases.},
  author={Goto, Masataka and Hashiguchi, Hiroki and Nishimura, Takuichi and Oka, Ryuichi},
  booktitle={ISMIR},
  volume={2},
  pages={287--288},
  year={2002}
}

@inproceedings{goto2006aist,
  title={AIST Annotation for the RWC Music Database.},
  author={Goto, Masataka and others},
  booktitle={ISMIR},
  pages={359--360},
  year={2006}
}

@inproceedings{driedger2019towards,
  title={Towards Automatically Correcting Tapped Beat Annotations for Music Recordings.},
  author={Driedger, Jonathan and Schreiber, Hendrik and De Haas, W Bas and M{\"u}ller, Meinard},
  booktitle={ISMIR},
  pages={200--207},
  year={2019}
}

@inproceedings{eremenko2018audio,
  title={Audio-aligned jazz harmony dataset for automatic chord transcription and corpus-based research},
  author={Eremenko, Vsevolod and Demirel, Emir and Bozkurt, Baris and Serra, Xavier},
  booktitle={Proceedings of the 19th International Society for Music Information Retrieval Conference},
  pages={483--490},
  address={Paris, France},
  year={2018}
}

@article{foster2021filosax,
  title={Filosax: A dataset of annotated jazz saxophone recordings},
  author={Foster, Dave and Dixon, Simon and others},
  year={2021}
}

@inproceedings{foscarin2020asap,
  title={ASAP: A dataset of aligned scores and performances for piano transcription},
  author={Foscarin, Francesco and Mcleod, Andrew and Rigaux, Philippe and Jacquemard, Florent and Sakai, Masahiko},
  booktitle={Proceedings of the 21st International Society for Music Information Retrieval Conference},
  pages={534--541},
  year={2020}
}

@inproceedings{gillick2019learning,
  title={Learning to groove with inverse sequence transformations},
  author={Gillick, Jon and Roberts, Adam and Engel, Jesse and Eck, Douglas and Bamman, David},
  booktitle={International conference on machine learning},
  pages={2269--2279},
  year={2019},
  organization={PMLR}
}

@inproceedings{xi2018guitarset,
  title={GuitarSet: A Dataset for Guitar Transcription.},
  author={Xi, Qingyang and Bittner, Rachel M and Pauwels, Johan and Ye, Xuzhou and Bello, Juan Pablo},
  year={2018},
  organization={Proceedings of the 19th International Society for Music Information Retrieval Conference}
}

@inproceedings{krebs2013rhythmic,
  title={Rhythmic Pattern Modeling for Beat and Downbeat Tracking in Musical Audio.},
  author={Krebs, Florian and B{\"o}ck, Sebastian and Widmer, Gerhard},
  booktitle={ISMIR},
  pages={227--232},
  year={2013}
}

@article{fujinaga2012one,
  title={ONE IN THE JUNGLE: Downbeat Detection in Hardcore, Jungle, and Drum and Bass},
  author={Fujinaga, Ichiro and Hockman, Jason and Davies, Matthew},
  year={2012}
}

@article{hainsworth2004particle,
  title={Particle filtering applied to musical tempo tracking},
  author={Hainsworth, Stephen W and Macleod, Malcolm D},
  journal={EURASIP Journal on Advances in Signal Processing},
  volume={2004},
  number={15},
  pages={927847},
  year={2004},
  publisher={Springer}
}

@book{gouyon2005computational,
  title={A computational approach to rhythm description-Audio features for the computation of rhythm periodicity functions and their use in tempo induction and music content processing},
  author={Gouyon, Fabien and others},
  year={2005},
  publisher={Universitat Pompeu Fabra}
}

@article{holzapfel2012selective,
  title={Selective sampling for beat tracking evaluation},
  author={Holzapfel, Andre and Davies, Matthew EP and Zapata, Jos{\'e} R and Oliveira, Jo{\~a}o Lobato and Gouyon, Fabien},
  journal={IEEE Transactions on Audio, Speech, and Language Processing},
  volume={20},
  number={9},
  pages={2539--2548},
  year={2012},
  publisher={IEEE}
}

@inproceedings{rocamora2015audio,
  title={An audio-visual database of Candombe performances for computational musicological studies},
  author={Rocamora, Mart{\'\i}n and Jure, Luis and Marenco, Bernardo and Fuentes, Magdalena and Lanzaro, Florencia and G{\'o}mez, Alvaro},
  booktitle={2do. Congreso Internacional de Ciencia y Tecnolog{\'\i}a Musical (CICTeM 2015), Buenos, Argentina, 17 y 19 de set, 2015},
  year={2015}
}

@inproceedings{marchand2015swing,
  title={Swing ratio estimation},
  author={Marchand, Ugo and Peeters, Geoffroy},
  booktitle={Digital Audio Effects 2015 (Dafx15)},
  year={2015}
}

@inproceedings{raffel2014mir_eval,
  title={MIR\_EVAL: A Transparent Implementation of Common MIR Metrics.},
  author={Raffel, Colin and McFee, Brian and Humphrey, Eric J and Salamon, Justin and Nieto, Oriol and Liang, Dawen and Ellis, Daniel PW and Raffel, C Colin},
  booktitle={ISMIR},
  volume={10},
  pages={2014},
  year={2014}
}

@article{loshchilov2017decoupled,
  title={Decoupled weight decay regularization},
  author={Loshchilov, Ilya and Hutter, Frank},
  journal={arXiv preprint arXiv:1711.05101},
  year={2017}
}

@inproceedings{bolt2026beat,
  title={Beat and Downbeat Detection: A Reformulated Approach},
  author={Bolt, James and Pauwels, Johan and Fazekas, George},
  booktitle={ICASSP 2026-2026 IEEE International Conference on Acoustics, Speech and Signal Processing (ICASSP)},
  pages={15282--15286},
  year={2026},
  organization={IEEE}
}

@inproceedings{ru2026toward,
  title={Toward Robust And Efficient Beat Tracking Via Beat-Aware Attention},
  author={Ru, Ganghui and Yu, Yi and Li, Wei},
  booktitle={ICASSP 2026-2026 IEEE International Conference on Acoustics, Speech and Signal Processing (ICASSP)},
  pages={15957--15961},
  year={2026},
  organization={IEEE}
}

@inproceedings{redmon2017yolo9000,
  title={YOLO9000: better, faster, stronger},
  author={Redmon, Joseph and Farhadi, Ali},
  booktitle={2017 IEEE conference on computer vision and pattern recognition (CVPR)},
  pages={6517--6525},
  year={2017},
  organization={Ieee}
}

\end{document}